\documentclass[english,a4paper,prl,twocolumn,amsmath,amssymb,longbibliography,superscriptaddress,nofootinbib]{revtex4-2}
\usepackage{graphicx,color}
\usepackage{physics}
\usepackage{siunitx}
\usepackage[colorlinks,linkcolor=magenta,citecolor=magenta]{hyperref}
\usepackage[english]{babel}

\newcommand{\kT}{k_\text{B}T}

\AtBeginDocument{%
    \newwrite\bibnotes
    \def\bibnotesext{Notes.bib}
    \immediate\openout\bibnotes=\jobname\bibnotesext
    \immediate\write\bibnotes{@CONTROL{REVTEX41Control}}
    \immediate\write\bibnotes{@CONTROL{%
    apsrev41Control,author="08",editor="1",pages="1",title="0",year="1"}}
     \if@filesw
     \immediate\write\@auxout{\string\citation{apsrev41Control}}%
    \fi
}%

\begin{document}
\title{Work Extraction via Backward Motion in Optimal Closed-Loop Stochastic Control}

\author{Luis Frieder Reinalter }
\affiliation{Fachbereich Physik, Universität Konstanz, 78464 Konstanz, Germany}
\author{Emanuele Panizon}
\affiliation{Area Science Park, Localit\`{a} Padriciano 99, 34149 Trieste, Italy}
\author{Lokesh Muruga }
\affiliation{Fachbereich Physik, Universität Konstanz, 78464 Konstanz, Germany}
\author{Clemens Bechinger}
\email[Corresponding author: ]{clemens.bechinger@uni-konstanz.de}
\affiliation{Fachbereich Physik, Universität Konstanz, 78464 Konstanz, Germany}
\affiliation{Centre for the Advanced Study of Collective Behaviour, University of Konstanz, Germany}

\begin{abstract}
We experimentally realize finite-time feedback control in an overdamped colloidal system using real-time optical tweezers with in situ reinforcement learning (RL). By varying the protocol duration \(t_f\) for displacing the optical trap between prescribed positions, the optimal strategies identified by RL reveal a crossover from deterministic dragging toward the target to feedback-assisted exploitation of thermal fluctuations, reducing and eventually overcoming the energetic cost. The resulting policies agree quantitatively with the exact optimal closed-loop solution. By extending the approach to spatially localized external forcing, we further show that RL can identify optimal feedback strategies in heterogeneous stochastic environments where direct analytical control design is challenging.
\end{abstract}

\maketitle

\emph{Introduction}---Optimal control aims to steer physical systems toward prescribed states while minimizing energetic costs~\cite{schmiedl_optimal_2007,aurell_optimal_2011,gingrich_near-optimal_2016,engel_optimal_2023,alvarado_optimal_2026, bechhoefer_control_2021,davis_optimal_2026}. At microscopic scales, however, thermal fluctuations strongly influence finite-time transport and fundamentally constrain how efficiently stochastic systems can be controlled. Stochastic thermodynamics has therefore motivated increasing interest in optimal finite-time protocols for fluctuating systems, including colloidal transport and information engines~\cite{seifert_stochastic_2012,sagawa_generalized_2010,toyabe_experimental_2010,saha_bayesian_2022,saha_information_2023,loos_universal_2024,PhysRevX.14.011012}.

While optimal finite-time open-loop control has been studied extensively in theory and experiment~\cite{schmiedl_optimal_2007,aurell_optimal_2011,loos_universal_2024,oikawa_experimentally_2025}, corresponding closed-loop strategies have only recently begun to attract attention. In particular, numerical work has shown that neural-network feedback controllers can learn demon-like protocols that extract work from fluctuating nanosystems~\cite{whitelam_demon_2023}. However, the experimental realization of optimal finite-time feedback protocols, and the microscopic structure of the resulting control actions, remain largely unexplored.

In open-loop control, driving protocols are optimized only regarding the average system response and therefore requires only statistical information about the dynamics. Feedback (closed-loop) control fundamentally changes this situation because each control action depends on the instantaneous state of the system obtained from real-time measurements. This allows for the exploitation of thermal fluctuations, enabling the controller to adjust its actions on to reduce energetic costs~\cite{abreu_extracting_2011,horowitz_designing_2011,admon_experimental_2018}.

These challenges become even more pronounced in the presence of additional external forces and spatially varying fluctuations, where analytical solutions are generally unavailable. Reinforcement learning~\cite{sutton_reinforcement_1998} offers a promising route to such problems because optimal policies can be learned directly from experimental trajectories without requiring a detailed dynamical model ~\cite{whitelam_demon_2023,xu_reinforcement_2022,nasiri_reinforcement_2022,rengifo_machine_2025}. 

Here, we experimentally realize optimal finite-time feedback control in an optical-tweezer setup with in situ reinforcement learning (RL). First, as an analytically solvable benchmark, we control the position of an optical trap containing a colloidal particle between prescribed initial and final trap positions within a fixed protocol duration \(t_f\), while minimizing the average work. By varying \(t_f\), the optimal feedback policies identified by RL reveal a crossover from a regime in which the particle is continuously pulled toward the target to one in which thermal fluctuations increasingly reduce the associated energetic cost and eventually lead to negative average work. The experimentally identified optimal policies are in quantitative agreement with the corresponding exact optimal closed-loop solution. Resolving individual feedback actions reveals that the reduction in energetic cost originates from backward trap displacements that extract work from the particle. Finally, we extend the approach to situations with additional external forces and spatially varying fluctuations, a regime where direct analytical control is challenging. Under such conditions, the RL-identified optimal strategy rapidly directs the particle toward the perturbed region, where externally enhanced fluctuations enable work extraction before the final displacement to the target is completed.

\emph{Experimental setup}---Our experiments are performed with a silica particle ($a = 1.37\,\mu\mathrm{m}$) confined by optical tweezers. The optical trap is generated by a 532\,nm laser focused through a $100\times$, 1.45-NA oil-immersion objective. Throughout the experiments, the trap stiffness is kept constant at $\kappa = 2\pm0.2\,\mathrm{pN}/\mu\mathrm{m}$, while only the trap position $\lambda_t$ is externally controlled. The particle is suspended in a water--glycerol mixture, which increases the viscous relaxation time in the trap to $\tau_r\approx 25\,\mathrm{ms}$, thereby facilitating real-time feedback control. Lateral trap steering is achieved with a piezo-controlled mirror in a $4f$-conjugate-plane configuration, yielding a positioning accuracy of $4\,\mathrm{nm}$ and an upper settling time of $1\,\mathrm{ms}$. Particle positions are measured with a spatial resolution of approximately $7\,\mathrm{nm}$ by digital video microscopy using a CMOS camera acquiring images every 12 $\mathrm{ms}$. The sample cell is kept constant at a temperature of 25$\pm 0.1\,^\circ$C (for details see End Matter and SM~\cite{noauthor_see_nodate}).

At each control step, separated by a time interval $\Delta t$, during the motion of the trap from an initial ($\lambda_i$) to a final position ($\lambda_f$) over a prescribed protocol duration $t_f$, the experimentally accessible state is defined by the measured particle position $x_t$, the instantaneous trap position $\lambda_t$, and the elapsed time $t$, $s_t=(x_t,\lambda_t,t)$.
Based on this state, the controller determines the subsequent trap position according to
$s_t \rightarrow \lambda_{t+\Delta t}$
such that the protocol is generated in real time from the measured stochastic trajectory. 
The work $W$ exerted on the particle during such protocols is determined according to ~\cite{sekimoto_langevin_1998,speck_distribution_2004}
\begin{align}
W
=
\int_0^{t_f}
\frac{\partial U(x_t,\lambda_t)}{\partial \lambda_t}
\dot{\lambda}_t\,\mathrm{d}t
=
-\kappa
\int_0^{t_f}
(x_t-\lambda_t)\dot{\lambda}_t\,\mathrm{d}t \label{Eq:Work} .
\end{align}
Within this definition, \(W>0\) is work exerted on the particle, whereas \(W<0\) is work performed by the particle.

To experimentally identify optimal feedback strategies, we employ a reinforcement-learning (RL) algorithm that communicates with the experimental setup in real time. The controller is represented by a fully connected artificial neural network (ANN)~\cite{mnih_human-level_2015}  and optimized using proximal policy optimization (PPO)~\cite{schulman2017proximal}. 
As input, the ANN receives the experimentally accessible state \(s_t\) and returns the next control action \(\lambda_{t+\Delta t}\), thereby defining the policy. 
During training, each control action is assigned the reward \(-\Delta W_t\).
The PPO algorithm updates the controller's parameters  using batches of stochastic realizations to reinforce policies that maximize the reward signals and thus minimize the average total work \(\langle W \rangle\) (see SM~\cite{noauthor_see_nodate}). To ensure the boundary condition defined by the prescribed final position \(\lambda_f\), deviations from the target at the end of the protocol are penalized by an additional terminal contribution to the reward function.

\begin{figure}
    \centering
    \includegraphics{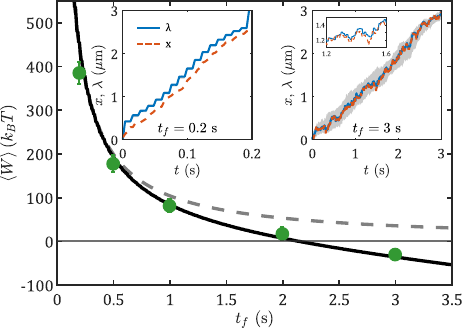}
    \caption{Average work \(\langle W\rangle\) versus the protocol duration \(t_f\). Symbols are experimental values obtained from RL-identified optimal policies, the solid and dashed lines denotes the analytical solution for the optimal closed-loop protocol and that without feedback. Insets: representative trap trajectories \(\lambda_t\) and corresponding particle trajectories \(x_t\) for \(t_f=\SI{0.2}{\second}\) and \(\SI{3}{\second}\). For \(t_f=\SI{0.2}{\second}\), the particle systematically lags behind the trap, and variations between experiments are barely visible. 
    For \(t_f=\SI{3}{\second}\), the particle remains closer to the trap trajectory, and individual realizations exhibit much larger scatter (gray band denotes scatter across 100 repeated realizations).}
    \label{fig:1}
\end{figure}

\begin{figure*}
    \centering
    \includegraphics[width=1\linewidth]{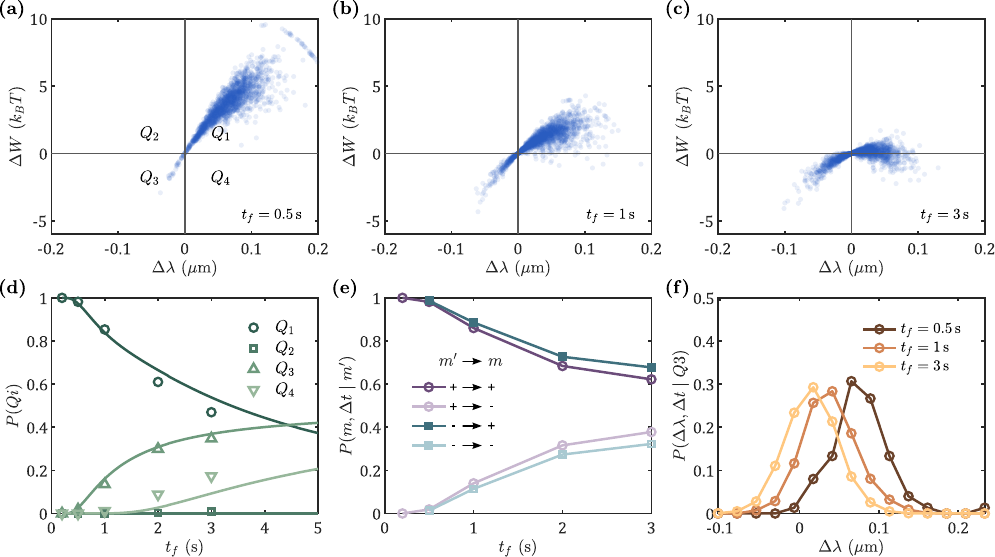}
    \caption{
(a)--(c) Representative action distributions for \(t_f=\SI{0.5}{\second}\), \(\SI{1}{\second}\), and \(\SI{3}{\second}\), respectively, classified by trap displacement \(\Delta\lambda\) and work increment \(\Delta W\). Forward and backward steps correspond to positive and negative \(\Delta\lambda\), respectively. The quadrants denote forward work-costing steps ($Q1$), backward work-costing steps ($Q2$), backward work-extracting steps ($Q3$), and forward work-extracting steps ($Q4$). Short protocols are dominated by $Q1$, whereas increasing \(t_f\) enhances the population of work-extracting $Q3$ events.
(d) Normalized quadrant occupation probabilities \(P(Qi)\) as a function of \(t_f\). Symbols are experimental data from RL-identified optimal policies; solid lines are theoretical predictions for the optimal closed-loop protocol.
(e) Conditional probabilities of consecutive trap displacement directions ${m',m}$. Forward and backward steps are denoted by \(+\) and \(-\), respectively.
(f) Distribution of the subsequent displacement \(\Delta\lambda_{t+\Delta t}\) after a $Q3$ event, showing that backward work-extracting steps are preferentially followed by compensating forward steps at short \(t_f\).
In (e) and (f), lines are linear connections between experimental data points and serve only as guides to the eye.
}
    \label{fig:Strategy}
\end{figure*}

\emph{Experimental results}---We first consider short protocol durations \(t_f\). Figure~\ref{fig:1} (left inset) shows a representative trap trajectory \(\lambda_t\) (solid blue) together with the corresponding particle trajectory \(x_t\) (dashed orange) after training for \(t_f=\SI{0.2}{\second}\). Throughout the protocol, \(x_t<\lambda_t\), indicating that the particle persistently lags behind the trap. The step-like structure of \(\lambda_t\) reflects the finite experimental feedback update time. Consistently, both \(\lambda_t\) and \(x_t\) exhibit only small variations between individual realizations; in the inset, the scatter of 100 realizations is hardly visible as a gray band. Notably, \(\lambda_t\) exhibits pronounced initial and final jumps, reminiscent of the boundary jumps known from optimal open-loop driving~\cite{schmiedl_optimal_2007,loos_universal_2024}. This similarity is expected because, for small \(t_f\), thermal fluctuations can hardly be exploited, causing the optimal feedback and open-loop strategies to become nearly identical.

For longer protocols, illustrated by $t_f = 3\,\mathrm{s}$, the feedback strategy changes qualitatively. The trap closely follows the particle trajectory, and both trajectories repeatedly cross, indicating that thermal fluctuations are exploited at the level of individual realizations rather than merely perturbing a deterministic dragging protocol. The larger scatter (gray band) between realizations further reflects the state-dependent character of the learned feedback.

The symbols in Fig.~\ref{fig:1} show the average work $\langle W \rangle$ obtained from experimentally identified optimal policies as a function of $t_f$. Each $\langle W \rangle$ is calculated from an ensemble of 100 trajectories. For short protocol durations, $\langle W \rangle$ is positive, demonstrating that energy must be supplied to move the trap to the target position. With increasing $t_f$, thermal fluctuations can be exploited more efficiently by the feedback controller, leading to a continuous decrease of $\langle W \rangle$. This effect becomes particularly pronounced for $t_f \gtrsim \SI{2}{\second}$, where $\langle W \rangle$ becomes negative, indicating that the control task is on average powered by thermal fluctuations of the equilibrium bath. The experimental data show excellent agreement with the corresponding values obtained from the analytical solution of the optimal closed-loop problem~\cite{panizon_optimal_2026} (see also End Matter), thereby validating the experimental RL approach used to identify the optimal feedback strategies.

For comparison, Fig.~\ref{fig:1} also shows the theoretically predicted optimal open-loop protocol (dashed line). As expected, for short $t_f$, the open- and closed-loop results coincide, reflecting the dominance of deterministic control. With increasing $t_f$, however, the two solutions progressively deviate. In particular, the open-loop protocol does not exhibit negative values of $\langle W \rangle$, demonstrating that without feedback control no net energy can be extracted from the equilibrium bath.

To understand how the RL-identified closed-loop strategy reduces the energetic cost, we resolve individual feedback actions by comparing each trap displacement $\Delta\lambda$ with the corresponding work increment $\Delta W$. Figures~2(a)--(c) show the resulting action distributions for increasing $t_f$. Short protocols are dominated by Q1 events, i.e., forward steps that perform work on the particle. With increasing $t_f$, the population of Q3 events grows, demonstrating that work reduction mainly arises from backward, work-extracting trap displacements rather than from favorable forward fluctuations.
\begin{figure*}
    \centering
    \includegraphics{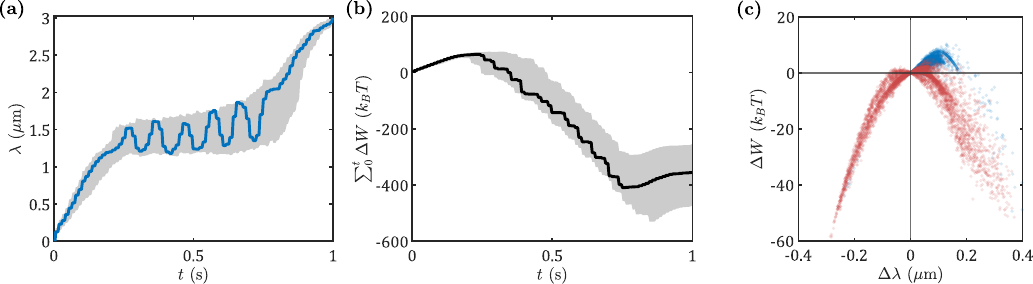}
    \caption{
(a) RL-optimized trap trajectory for \(t_f=\SI{1}{\second}\) in the presence of an externally imposed, time-dependent perturbation. The blue curve shows a single realization of the optimal feedback protocol identified by RL, while the gray band indicates the scatter between 100 individual experiments. The trap trajectory first approaches the perturbed region, remains there for an extended part of the protocol, and then reaches the target. Oscillations visible in individual realizations arise from the external drive and largely average out over many realizations because the driving phase is randomized.
(b) Corresponding accumulated work for the same protocol. After an initial increase, the accumulated work decreases over an extended time interval before the final approach to the target.
(c) Work increments \(\Delta W\) versus trap displacements \(\Delta\lambda\). Points are color-coded according to whether the particle is inside the perturbation region, defined by the width \(\sigma\) (red), or outside it (blue). Negative work increments inside the perturbed region occur for both backward and forward trap displacements, corresponding to $Q3$ and $Q4$, respectively.
}
    \label{fig:Perturbed}
\end{figure*}
Figure~\ref{fig:Strategy}(d) quantifies this redistribution in terms of the normalized quadrant occupation probabilities \(P(Qi)\). With increasing \(t_f\), \(P(Q1)\) decreases, whereas \(P(Q3)\) increases and becomes the dominant work-extracting contribution in the experimentally relevant range. The remaining quadrants play only a minor role: \(P(Q2)\) stays close to zero, indicating that energetically unfavorable backward steps are strongly suppressed, while \(P(Q4)\), corresponding to forward work-extracting steps, remains comparatively small. The feedback controller therefore does not mainly wait for favorable forward fluctuations. Rather, it actively exploits fluctuations by transiently moving the trap backward to recover work from the particle. The solid lines show the theoretical predictions (see End Matter) for the steady-state limit of the
optimal closed-loop protocol in the equilibrium bath, which agree well 
with the experimental data from RL-identified optimal policies.

As shown in Fig.~2(e), consecutive trap displacements are classified by their directions, with $+$ and $-$ denoting forward and backward steps; the four possible pairs are therefore $++$, $+-$, $-+$, and $--$. Consecutive forward steps dominate at short $t_f$, whereas backward--forward sequences become relevant once Q3 events appear. Figure~2(f) confirms that, especially for short protocols, backward work-extracting steps are often followed by compensating forward motion. At longer $t_f$, this bias weakens, consistent with more flexible feedback-controlled trajectories.

\emph{Feedback control with external forces}---So far, we have considered feedback control in an equilibrium environment, where the controller can only exploit thermal fluctuations. In many  situations, however, additional external flows or forces drive the environment out of equilibrium and thereby modify the optimal feedback strategy. We therefore next consider feedback control for \(t_f=\SI{1}{\second}\) in the presence of externally imposed, spatially localized and time-dependent forces. As before, the optimal strategy is identified from experimental data using RL.

Experimentally, external forcing is realized by moving the sample cell relative to the optical trap with a piezo-driven stage. In the reference frame of the trap, this relative motion generates a viscous Stokes force on the particle in addition to the thermal fluctuations. To prevent the controller from simply adapting to a fixed deterministic drive, the phase \(\Phi\) of the imposed stage motion is randomized between realizations. Moreover, the perturbation is spatially localized, creating a region with enhanced nonequilibrium fluctuations within an otherwise unperturbed control landscape. 
The controller must therefore learn not only how to move toward the target, but also whether, when, and for how long it is beneficial to visit this perturbed region in order to reduce the total work. The imposed stage motion is given by
\[
x_{\mathrm{stage}}(t,x)
=
A_0
\exp\left[
-\frac{(x-x_c)^2}{2\sigma^2}
\right]
\sin(2\pi f t + \Phi),
\]
where \(A_0=\SI{2.5}{\micro\meter}\) and \(f=\SI{10}{\Hz}\) denote the driving amplitude and frequency, respectively. The modulation depends on the spatial coordinate \(x\), is centered in the middle of the control region at \(x_c=\SI{1.5}{\micro\meter}\), and has a width \(\sigma=\SI{400}{\nano\meter}\).

Figure~\ref{fig:Perturbed}(a) shows the optimal trap dynamics identified by RL in this heterogeneous nonequilibrium environment. The solid blue line denotes a single realization of \(\lambda_t\), while the gray band indicates the full scatter across 100 realizations.
The controller first steers the particle toward the perturbed region, where it remains for a substantial fraction of the protocol before completing the task. The single realization exhibits pronounced oscillatory motion induced by the external drive, while these oscillations largely average out in \(\langle\lambda_t\rangle\) because of the randomization of the phase \(\Phi\) between realizations (gray band). This illustrates how the feedback strategy, unlike the equilibrium case, adapts to the spatially localized perturbation.

Figure~\ref{fig:Perturbed}(b) shows the time-dependent accumulated work for a single realization (black) and the full scatter over 100 experiments (gray). The accumulated work initially increases as the controller steers the particle toward the perturbed region, but then decreases over an extended part of the protocol before the trap is finally moved to the target. Comparing Fig.~\ref{fig:Perturbed}(b) with Fig.~\ref{fig:Perturbed}(a) shows that this decrease occurs on the same time scale as the oscillatory trap motion, indicating that work is extracted while the particle explores the perturbed region. This interpretation is confirmed microscopically in Fig.~\ref{fig:Perturbed}(c), where work increments are plotted against the corresponding trap displacements in the same \(\Delta W\)--\(\Delta\lambda\) representation introduced above. A binary symbol color code distinguishes particle positions inside the perturbation region, defined by the width \(\sigma\) (red), from those outside this region (blue). Within the perturbed region, large negative work increments occur for both backward and forward steps, corresponding to $Q3$ and $Q4$, respectively, whereas outside this region the work increments are predominantly positive and mainly correspond to forward work-costing steps in $Q1$.


\textit{Summary---}Our experiments demonstrate optimal finite-time feedback control in stochastic systems using real-time reinforcement learning. In equilibrium environments, the learned policies reveal a crossover from deterministic dragging to fluctuation-assisted work extraction, in quantitative agreement with the exact optimal closed-loop solution. Resolving individual actions shows that this reduction originates from state-dependent backward trap displacements. In heterogeneous nonequilibrium environments, RL further identifies strategies that exploit spatially localized external forcing, establishing it as a versatile route to experimental optimal feedback control beyond analytically tractable settings.


\emph{Acknowledgments}---CB acknowledges funding by the ERC (Grant No. 101141477). LM acknowledges support by the Alexander von Humboldt foundation. We thank S. Monter, V.-L. Heuthe, and F. Ginot for fruitful discussions. The authors declare no competing interests.

\emph{Acknowledgments of AI use}---The authors used Gemini, Claude, and ChatGPT for Python-coding assistance and text polishing, and take full responsibility for the manuscript.


\emph{Data availability}---The data that support the findings of this Letter are openly available~\cite{data}.


\bibliography{references.bib}

\onecolumngrid
\section*{End Matter}
\twocolumngrid
\renewcommand{\theequation}{A\arabic{equation}}
\setcounter{equation}{0}

\emph{Experimental apparatus}---The optical trap was generated by a laser with wavelength \(\lambda_{\mathrm{trap}}=\SI{532}{\nano\meter}\), focused through a \(100\times\), \(1.45\)-NA oil-immersion objective (Olympus). Lateral control of the trap position \(\lambda_t\) was achieved with a piezo-controlled steering mirror (Piezo Concepts TT2) placed in a \(4f\)-conjugate plane of the microscope, yielding a positioning accuracy of approximately \(\SI{4}{\nano\meter}\) and an upper settling time of approximately \(\SI{1}{\milli\second}\). The steering mirror was driven by analog output signals generated by a control computer and a data-acquisition device (DAQ, NI USB-6003).

Particle tracking was performed in an inverted-microscope configuration using \(\SI{450}{\nano\meter}\) LED illumination and a CMOS camera (Basler acA800-510mu), acquiring images every \(\SI{12}{\milli\second}\). The particle center was extracted online from each camera frame and used, together with the current trap position and protocol time, to compute the next trap position. Camera exposure, image analysis, DAQ output, and trap steering were synchronized such that each update was completed within one feedback cycle, with a latency between image acquisition and trap update of approximately \(\SI{1.6}{\milli\second}\).

Samples consisted of silica particles (MicroParticles) with radius \(a=\SI{1.37}{\micro\meter}\), suspended in an \(8{:}2\) water--glycerol mixture and loaded into glass capillaries of height \(\SI{100}{\micro\meter}\), which were sealed with beeswax and epoxy resin. The trap stiffness \(\kappa\) was calibrated from the equilibrium particle distribution \(p(x)\propto\exp[-U(x)/(k_{\mathrm{B}}T)]\), using the harmonic potential \(U(x)=\kappa(x-\lambda)^2/2\), yielding \(\kappa=2\pm0.2\,\mathrm{pN}/\mu\mathrm{m}\). The friction coefficient \(\gamma\) was independently obtained from fits to the short-time mean-squared displacement, giving a relaxation time \(\tau_r=\gamma/\kappa\approx\SI{25}{\milli\second}\)~\cite{jones_optical_2015,data}.

External forcing, when applied, was generated by translating the sample cell with a piezo-actuated stage (Piezo Concepts LT300). During all measurements, the sample holder and objective were actively temperature-stabilized at \(25\pm0.1\,^\circ\mathrm{C}\) using resistive heating elements (Okolab and Thorlabs) mounted in direct thermal contact with the sample chamber. The trapping laser temperature was stabilized by a water-cooled thermostat to improve laser stability. The entire setup was enclosed, limiting apparatus temperature variations to approximately \(\pm\SI{0.5}{\degreeCelsius}\), while the room temperature was maintained at \(22\pm1\,^\circ\mathrm{C}\).

\begin{figure}
    \centering
    \includegraphics[width=1\linewidth]{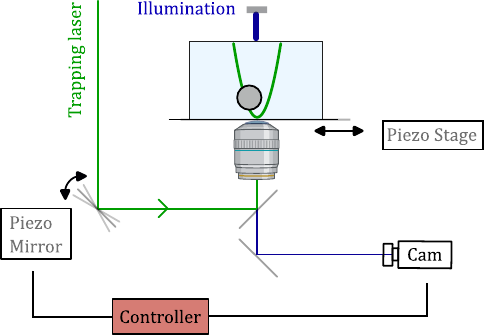}
    \caption{
Schematic of the experimental optical-tweezer setup. 
A silica colloid is confined in an optical trap generated by a focused trapping laser. The trap position is controlled by a piezo-actuated steering mirror, while the particle position is recorded by video microscopy. The measured particle position, current trap position, and protocol time are passed to a real-time controller, which computes the next trap displacement. External perturbations can be applied by translating the sample with a piezo stage.
}
    \label{fig:setup}
\end{figure}

\emph{Optimal work in equilibrium baths}---When the bath is in equilibrium, the expected work under optimal control can be evaluated in closed form.
In the Linear-Quadratic-Gaussian (LQG) formulation~\cite{bechhoefer_control_2021,panizon_optimal_2026}, this is equivalent to computing the Value Function at the initial state of a particle at equilibrium, with a distribution centered around $\mu=0$ and a dispersion of $\kT/\kappa$, and initial trap position $\lambda_0=0$. Following the convention in ~\cite{panizon_optimal_2026}, we consider $N$ as the total number of steps, and $n$ as the remaining steps from $t_f$.
The value $V_N(\mu=0, \lambda=0, \Sigma=\kT/\kappa)$ decouples into independent spatial, $J_N(\mu=0,\lambda=0)$, and informational, $g_N(\Sigma=\kT/\kappa)$, costs.
The first term corresponds to the deterministic cost of steering:

\begin{equation}
    J_N(0, 0) = \frac{1}{2} (P_N+\kappa) \lambda_f^2.
\end{equation}

Here, $P_N$ is the Riccati term $P_n = -\kappa \left[ \frac{n(1-\alpha)}{1+\alpha+n(1-\alpha)} \right]$~\cite{panizon_optimal_2026} evaluated at $n=N$, and $\alpha = \exp(-\kappa\Delta t/\gamma)$. This term recovers the known open-loop Schmiedl-Seifert result $\langle W_{ol} \rangle = \frac{\kappa\lambda_f^2}{2 + t_f/\tau_r}$ in the $\Delta t \rightarrow0$ limit.

The second term $g_N(\Sigma)$ needs to be solved recursively, noting that the variance varies deterministically during a trajectory. It starts at $\Sigma_N=\kT/\kappa$ and, while at each step it is set to zero by the observation, it always grows by thermal diffusion to $\Sigma_{eq}=\frac{k_B T}{\kappa}(1-\alpha^2)$. The informational cost can then be evaluated by the recurrence ~\cite{panizon_optimal_2026} $g_n(\Sigma) = \frac{1}{2}P_n\Sigma + g_{n-1}(\Sigma_{eq})$. Unrolling this sequence using the known values for $\Sigma_n$ and $P_n$ gives:
\begin{equation}
\begin{split}
    g_N(\kT/\kappa) = \\
    \frac{1}{2} P_N \kT/\kappa - \frac{1}{2}\Sigma_{eq}\kappa \Big( N - \nu \big[ \psi(N + \nu) - \psi(\nu) \big] \Big),
\end{split}
\end{equation}

where we have used the digamma function identity $\sum_{n=0}^{N-1} (\nu + n)^{-1} = \psi(N + \nu) - \psi(\nu)$ for $\nu = (1+\alpha)/(1-\alpha)$.

Combining these two quantities allows for the following closed-form formulation for the average work:
\begin{equation}
\begin{split}
    \langle W \rangle = \underbrace{ \frac{1}{2} (P_N+\kappa) \lambda_f^2 }_{\text{Steering}} \\
    + \underbrace{ \frac{1}{2} P_N \left( \frac{k_B T}{\kappa} \right) - \frac{1}{2} \Sigma_{eq} \kappa \Big( N - \nu \big[ \psi(N + \nu) - \psi(\nu) \big] \Big) }_{\text{Fluctuations}}.
\end{split}
\label{eq:closed_loop_work}
\end{equation}

\emph{Quadrant occupancy}---To evaluate the probability that an instantaneous jump falls into a quadrant in the \(\Delta\lambda\) - \(\Delta W\) plane, it is convenient to factorize the  
work as $W_t = -\frac{1}{2}\kappa \Delta \lambda_t Z_t$ where $\Delta \lambda_t = \lambda_t - \lambda_{t-\Delta t}$ is the trap translation and $Z_t = 2x_t - \lambda_t - \lambda_{t-\Delta t}$ is the spatial offset. The relevant quadrant is defined by the signs of these two quantities. For a bath in equilibrium, $\Delta \lambda_t$ and $Z_t$ are jointly Gaussian. 

For a finite-time trajectory, their distributions depend on the instantaneous time $t$, meaning that the average quadrant occupations vary dynamically. However, we can define a non-equilibrium-steady-state (NESS) limit, letting $\lambda_f, t_f \rightarrow \infty$ while keeping the macroscopic velocity $v = \lambda_f / t_f$ constant. In this limit, the optimal control tracks the particle instantaneously $\lambda_t = x_t + v \Delta t \left( \frac{1}{1-\alpha} \right)$\cite{panizon_optimal_2026}. The distributions become time-independent, and it is possible to compute their means in terms of physical parameters only:
$\mu_{\Delta\lambda} =v \Delta t$, $\mu_Z = -v \Delta t (1+\alpha)/(1-\alpha)$.

The variances depend only on the thermal diffusion accumulated in one step: $\Sigma_{eq} = \frac{k_B T}{\kappa}(1-\alpha^2)$, and yield simply $\sigma_{\Delta\lambda}^2 = \sigma_Z^2 = \text{Cov}(\Delta, Z) = \Sigma_{eq}$.

Defining the normalized means $h = \mu_{\Delta\lambda}/\sqrt{\Sigma_{eq}}$ and $k = \mu_Z/\sqrt{\Sigma_{eq}}$, and their Pearson correlation $\rho=1$, one can express the probability of occupying a quadrant integrating the regions with the corresponding signs of $\Delta$ and $Z$. Using the Bivariate Normal Cumulative Distribution Function, $\Phi_2(x,y,\rho)$, this reads:
\begin{align}
    P(Q_1) &= \mathrm{\Phi}_2(h, -k, -1), \quad P(Q_2) = \mathrm{\Phi}_2(-h, k, -1), \\
    P(Q_3) &= \Phi_2(-h, -k, 1), \quad P(Q_4) = \Phi_2(h, k, 1).
\end{align}

Although exact only in the NESS limit, these approximations hold remarkably well over the full trajectories when the initial and final jumps are excluded.

\end{document}